\documentclass[11pt]{article}

\usepackage[final]{acl}

\usepackage{times}
\usepackage{latexsym}
\usepackage{booktabs,tabularx}

\usepackage[T1]{fontenc}

\usepackage[utf8]{inputenc}

\usepackage{microtype}

\usepackage{inconsolata}

\usepackage{graphicx}

\title{The Surge of Anti-Semitism in German Social Media \\ following the October 7 Attacks}

\author{Gregor Wiedemann \and Daniel Wehrend \\
  Leibniz-Institute for Media Research | Hans-Bredow-Institut \\
  Hamburg, Germany \\
  \texttt{\{g.wiedemann, d.wehrend\}@leibniz-hbi.de} 
}

\begin{document}
\maketitle
\begin{abstract}
We investigate the extent to which the Hamas attacks on Israel of October 7, 2023, have affected German social media debates about Judaism and Israel. For this, we develop an approach to detect 26 anti-Semitic categories in user postings via large language models (LLMs). The approach is applied to Facebook and Telegram posts ($N=125,718$) from three months before and after the event.
Methodically, we test different open-weight models in two setups---with and without user information as additional context to the post text. 
The best setup achieves up to 83~\% F1-score for binary anti-Semitism detection on our manually coded validation set. User context provides valuable information for most LLMs and drastically reduces false positives, for example, when (critically) reporting on anti-Semitic incidents.
Concerning our topic, we find that anti-Semitism is surging significantly on both platforms, while being about ten times more prevalent on Telegram compared to Facebook. Facebook users express anti-Semitic views most likely in posts about an alleged genocide in Gaza carried out by the Israeli army, whereas classic anti-Semitic stereotypes related to power and conspiracy theories are dominant on Telegram. After the attack, the discourse patterns on both platforms show signs of convergence, as classic anti-Semitism increases on Facebook, whereas Israel-related categories surge on Telegram.
\end{abstract}

\section{Introduction}

Anti-Semitism has been part of online communication since the early public internet era \cite{Conway2019Right-wing}. News events related to Judaism or the State of Israel often lead to intensified online debates \cite{Salminen2020Topic-driven}, in the course of which anti-Semitic statements increase, too.
On October 7, 2023, Palestinian Hamas militants, along with a large number of civilian participants from Gaza, carried out an attack on Israel that resulted in the deaths of around 1,200 people, left more than 5,000 others injured, and about 250 kidnapped and held hostage in the Gaza Strip \cite{andreev_arabic_2024}. Israel reacted with a military offensive against Gaza that killed tens of thousands of Palestinians over the months and years after the attack.
Social media users worldwide commented on the attack and its aftermath in the wider context of the Middle East conflict. Jewish people and the State of Israel have increasingly been targets of anti-Semitism in online communication in recent years \cite{adf_quantifying_2018}. At the same time, for historical reasons, the German public debate is very sympathetic to Israel's fate. Against this background, we are interested in determining to what extent the attack on October 7 has led to further polarization and, as a result, to a rise in anti-Semitism in German online debates.

To study the spread and characteristics of anti-Semitism in German social media communication, we investigate public user posts related to Judaism or the state of Israel from Facebook and Telegram during a period of three months before and after the attack. 
To study the discourse development quantitatively at a large scale, we ask three research questions: 
\begin{enumerate}
    \item To what extent can antisemitism—and in particular its specific manifestations, such as Holocaust denial, conspiracy theories, or the denial of Israel’s right to exist—be automatically detected in German-language social media posts?
    \item Does providing user information such as names and account handles as additional context to the post text improve the automatic classification of antisemitism with LLMs?
    \item Based on aggregate measures of classifier labels, how did the October 7 event affect the prevalence and salience of different categories of antisemitism on both platforms? 
\end{enumerate}
To answer these questions, we approach the challenge of operationalizing a complex academic definition of antisemitism (see Section~\ref{sec:rel_works}) through automated content analysis using large language models (LLMs). We apply LLM-based text classification on a large German language dataset containing (near) full samples of posts from both platforms that match topic-specific keywords. For model selection and prompt optimization, we manually annotated a validation sample of about 800 posts with a domain expert from antisemitism research. The predictions of the final model on the full dataset form the basis of our quantitative study. Accordingly, there are three main contributions of the paper:
\begin{itemize}
    \item We evaluate the performance of state-of-the-art open-weight LLMs for the detection of a complex set of anti-Semitic categories in German-language texts. Our results demonstrate that the best-performing models can achieve satisfactory results for further quantitative analysis.
    \item Providing additional context information about the source of a post to the LLM effectively reduces false positive classifications where, for instance, users criticize antisemitic statements or newspapers report neutrally about anti-Semitic events---a frequent source of error as reported by previous studies \cite{patel_evaluating_2025}. 
    \item Based on a quantitative, automated content analysis, we identify significant differences in the prevalence of anti-Semitism on Facebook and Telegram. In the aftermath of October 7, however, we observe a convergence in the use of anti-Semitic language patterns on both social media platforms, which represent mainstream and fringe voices in the German public discourse.
\end{itemize}
In the following section, we start with our definition of anti-Semitism in light of the academic discourse and position our study in the field of previous approaches to automatic anti-Semitism detection and social media reactions to the October 7 attack. Section~\ref{sec:data} reports on our approach to data collection, category system development, as well as manual expert and automatic coding. Section~\ref{sec:results} presents the results answering our three research questions. In the final Section, we summarize our findings and discuss the limitations of our approach to studying the spread of anti-Semitism online. 

\section{Related work}
\label{sec:rel_works}

In the academic discourse, two particularly influential, yet contested, approaches to defining antisemitism currently dominate: the \textit{International Holocaust Remembrance Alliance} (IHRA) working definition and the \textit{Jerusalem Declaration on Antisemitism} (JDA). Both definitions are the subject of controversial debate, in part because of their practical applicability and their alleged (negative) impact on freedom of expression and academic freedom. 

\subsection{Definitions of Anti-Semitism}

The IHRA definition conceptualizes anti-Semitism as a specific perception of Jews that may be expressed as hatred toward Jews, and that can target Jewish (and non-Jewish) individuals and their property as well as the Jewish community or religious institutions.\footnote{\url{https://holocaustremembrance.com/resources/working-definition-antisemitism}} It is accompanied by illustrative examples, including several about Israel-related antisemitism, some of which are frequently criticized for blurring the boundary between anti-Semitic hostility and legitimate criticism of Israeli state policy \cite{ruth_gould_ihra_2020}. However, the IHRA definition is widely used in research and institutional practice.

The JDA, in contrast, defines antisemitism as ``discrimination, prejudice, hostility or violence against Jews as Jews (or Jewish institutions as Jewish)''\footnote{\url{https://jerusalemdeclaration.org/}} and distinguishes (1) ``classic'' antisemitism, (2) Israel-related antisemitism, and (3) critical statements that are not per se antisemitic, explicitly stressing context-sensitivity and borderline cases. Particularly, this context-sensitivity makes the JDA a difficult framework for large-scale computational classification, because many of its distinctions require interpretive judgments about context, pragmatic meaning, and speaker intent that are usually not observable in text data.

Building on this debate, the present study adopts the IHRA definition as its overarching definitional baseline, but implements a project-specific operationalization through 26 sub-categories that are specified in a dedicated codebook, including diagnostic indicators as well as positive and negative examples for systematic boundary-setting. In addition, we investigate whether providing speaker information as additional context, in conjunction with the world knowledge encoded in LLMs, can improve the automatic detection of antisemitism.

\subsection{Automatic Anti-Semitism detection}

Automatic hate speech detection has become a mature subfield of NLP and computational social science. \citet{schmidt_survey_2017} and \citet{malik_deep_2025} provide comprehensive surveys that synthesize the methodological landscape and recurring challenges such as concept ambiguity, annotation subjectivity, and domain shift. Hate speech detection provides a baseline for more specific hate phenomena such as antisemitism. However, antisemitism has been addressed far less frequently in NLP. Existing antisemitism-focused work highlights the importance of high-quality annotation guidance based on widely acknowledged definitions and careful treatment of borderline cases (e.g., reporting/quoting antisemitism versus expressing it) \citep{jikeli_detecting_2021,jikeli_antisemitic_2023,steffen_codes_2023}.

Following the Hamas attacks on October 7, 2023, a substantial share of systematic monitoring and reporting on antisemitism online has been produced by civil-society organizations. Most of these works look qualitatively at small samples of data from social media platforms. For example, \citet{andreev_arabic_2024} identify a shift in the Arabic discourse from promoting traditional stereotypes of Jews to direct violence against the Jewish people. 
\citet{rose_narratives_2024} observe that the anti-Semitic surge of social media posts in social media is accompanied by an increase in anti-Muslim posts. The German discourse has been qualitatively examined by \citet{becker_celebrating_2023,becker_decoding_2024b} who, among other things, identified ``a turning point in antisemitic online communication, characterized by open celebration and affirmation of violence inflicted on Israeli civilians'' (p.~6).

Public antisemitism-specific language resources of academic origin remain limited. A prominent contribution is an English-language Twitter dataset with detailed annotation guidance, explicitly grounded in the IHRA definition \citep{jikeli_antisemitic_2023}. It is designed to reduce false positives by separating antisemitic content from merely related discourse (e.g., Holocaust remembrance, news reporting). For German, a labeled dataset and annotation guide for online antisemitism in the context of the COVID-19 pandemic was published by \citet{steffen_codes_2023} as a domain-specific resource.
Domain-agnostic datasets are English-only and, thus, cannot be transferred to German antisemitism detection without substantial loss from cross-lingual and domain shift. 

Existing computational approaches largely follow the broader hate-speech pipeline (supervised classifiers and transformer-based models) \cite{wiedemann-etal-2020-uhh} but must address antisemitism-specific challenges such as implicitness, quotation/reporting, and the dependence on fine-grained definitions. Prior work reports practical difficulties of robust automatic classification in German online environments and emphasizes the need for careful methodological design and oversight \citep{steffen_codes_2023}.

Most relevant to the present study, \citet{patel_evaluating_2025} evaluate multiple open-source LLMs for detecting antisemitism using in-context policy guidance (IHRA) and prompting variants, finding meaningful performance differences and utility issues across models. However, their setup operationalizes the task primarily as a binary decision (antisemitic vs.\ non-antisemitic), leaving open how well LLMs can support more fine-grained typologies required for substantive content analysis.

Building on this landscape, our study combines scalable LLM-based classification with a project-specific, fine-grained operationalization of antisemitism tailored to German-language social media and anchored in a dedicated content-analysis codebook. This enables analyses that go beyond binary detection to differentiate forms of antisemitism and to separate antisemitic content from semantically adjacent but non-antisemitic discourse.

\section{Data and Methodology}
\label{sec:data}

To investigate the spread and character of anti-Semitism in German social media, we collected a large dataset, developed an extensive codebook, annotated and validated a validation sample, utilized this to determine the performance of various LLMs, and applied one final, most efficient setup to detect anti-Semitic posts in the overall dataset.

\subsection{Data Collection}
We collected data from two social media platforms, Facebook and Telegram, with the help of keyword lists. 
With about 23 million weekly users, Facebook is the second most important social media platform in Germany in 2023, just behind Instagram (24 million weekly users) \cite{tippelt_social_2025}. With its text-focused content, it is an important debate forum for the German public discourse. 
With about 13 million weekly users \cite{koch_reichweiten_2022}, Telegram is less important in terms of the widespread use of its application. However, with its unique affordances for unmoderated communication to very large audiences, the platform became popular for far-right political communication, particularly among people who experienced restrictions on their expressed opinions on other leading platforms \cite{wiedemann_telegram_2023}. We therefore included Telegram in this study to compare Facebook as a mainstream platform with a fringe platform on which a higher proportion of anti-Semitic posts are likely to be found.

Data from Facebook was collected via the service CrowdTangle (CT) that was operated by Meta until August 2024, providing API-based access for researchers and journalists to `organic' posts (excl. ads, direct messages or comments) from public pages and groups above a certain (undisclosed) prominence threshold. We collected all posts between July 7, 2023, and January 7, 2024, i.e., three months before and after the October 7 attack, that were written in the German language and matched at least one of the following keywords:
\textit{Jude, jüdisch, anti-jüdisch, antijüdisch, Israel, israelisch,anti-israelisch, antiisraelisch, Zion, zionistisch, anti-zionistisch, antizionistisch}, and all of their inflected forms, as CT fulltext search does not perform any stemming of query terms. 

The keywords were selected with the goal of describing a base population of posts from the public German social media discourse that refer to Jewish or Israeli affairs. 
However, it is important to note that this approach may not capture all forms of anti-Semitic speech.
To clean the query result, we removed posts referencing popular person names such as Jude Bellingham, Jude Law, etc., from the dataset.

For the Telegram dataset, we relied on a very large collection of German Telegram posts compiled by the project Social Media Observatory (SMO). The SMO collects, among other things, social media communication by German public speakers \cite{wiedemann_concept_2023}. For Telegram, a large number of German-language channels were surveyed in 2023 using snowball sampling, and their public posts were then collected. From about 15,000 channels, 1,001 distributed content that matched our keyword list during the study period (see Tab.~\ref{tab:metrics_by_platform}). Again, we removed false-positive posts referencing first names.

\begin{table}[]
\resizebox{.485\textwidth}{!}{%
\begin{tabular}{@{}lrr@{}}
\toprule
\textbf{Platform}           & \textbf{Facebook} & \textbf{Telegram} \\ \midrule
N posts                     & 58,411            & 67,307            \\
Share pre Oct. 7            & 21.2 \%           & 9.6 \%            \\
Share post Oct. 7           & 78.8 \%           & 90.4 \%           \\
Avg. text length (chars)    & 622               & 649               \\
N users/channels            & 12,577            & 1,000             \\
Avg. posts per user/channel & 4.64              & 67.31             \\ \bottomrule
\end{tabular}%
}
\caption{Dataset statistics by platform}
\label{tab:metrics_by_platform}
\end{table}

As we retrieved data via thematically non-specific search terms, we consider the collection to be an almost complete sample for the public German-language discourse about affairs related to Judaism and/or Israel from both platforms (see Tab.~\ref{tab:metrics_by_platform} for dataset statistics).\footnote{With the limitation on public content made available via CT and the coverage of German-language channels across the entire Telegram-sphere in the SMO collection.} 
While data for both platforms is of comparable size in terms of the number of posts and average text length, we see significant differences in the increase of posts after October 7, and the number of users/channels and their activity in terms of posts per user/channel. The latter differences stem from the specific affordance of one-to-many communication in Telegram channels, where channel owners continuously speak to their own partial public. 

\begin{figure}
    \centering
    \includegraphics[width=1\linewidth]{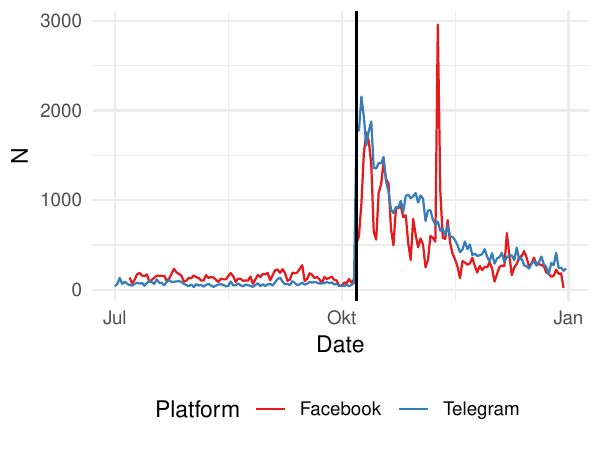}
    \caption{Post frequency per platform over time. The vertical line marks October 7, 2023.}
    \label{fig:timeseries}
\end{figure}

Fig.~\ref{fig:timeseries} shows the distribution of the number of posts on both platforms over time. The striking increase, together with the switch of the dominant source from Facebook to Telegram since October 7, reinforces our research hypothesis about the possible rise in anti-Semitic posts. The second peak for Facebook can be traced back to the German memorial day for the Reichspogromnacht, which is commemorated annually on November 9 and attracted particular attention after the Hamas attack.

\subsection{Data Annotation}

In adaptation of the IHRA definition of anti-Semitism and based on previous works in the project `Decoding Antisemitism' \cite{becker_decoding_2022,becker_decoding_2023,becker_decoding_2024b}, we developed a codebook consisting of 26 categories describing different types of anti-Semitic communication (see Tab.~\ref{tab:codebook}).
As the individual categories are distributed very unevenly and, in many cases, occur only rarely within the dataset, we utilized an iterated approach of zero-shot and few-shot in-context learning with a selection of open-weight LLMs to sample data for the validation set for human expert coding. In the first round of sampling, we started with the codebook definition as a system prompt and a sample of 10,000 posts in a zero-shot setting to generate label candidates. 

From the results, we sampled 100 examples per platform, one half randomly and the other half as a stratified, balanced sample for all categories. Afterwards, the sample was manually coded by a domain expert.\footnote{The expert was a master's student in the “Interdisciplinary Research on Antisemitism” program at the Center for Research on Antisemitism (ZfA) at the Technical University of Berlin, close to completing his degree.} 
Selected examples from the manual annotation were used in further rounds to extend the codebook and transform the zero-shot into a few-shot setting. This sampling process was repeated for four rounds with a different LLM for each round, whereas the last round was not stratified for platform balance. It resulted in a validation set with 803 posts, 335 from Facebook and 468 from Telegram. The combination of random and stratified sampling, as well as the use of different LLMs during the sampling rounds, was employed to achieve two conflicting goals: covering a representative range of characteristics of the complete dataset while also giving appropriate consideration to small categories.

Finally, we tested inter-coder reliability on the binary and subcategory level with a second, non-expert coder\footnote{The non-expert coder was a master's student from computer linguistics close to his degree.} on a random 50~\% sample of the validation set. For 400 binary label decisions, the two coders achieved a substantial Cohen's Kappa agreement of 71.5~\% (ca. 86~\% agreeing labels). Table~\ref{tab:codebook} shows the category-wise Kappa-agreement values and category sizes after expert labeling. Despite the stratification for category balance, for some categories, only a few examples could be retrieved. Partly due to the small sample sizes, substantial or moderate agreement was achieved in only 16 of the 26 subcategories ($\kappa>0.4$). For our ground truth, we decided to rely on the expert labels. But analysis results for the small categories must be viewed with particular caution.

\subsection{Model selection}

\begin{table*}[]
\centering
{\fontsize{9.5}{11.5}\selectfont
\begin{tabular}{@{}lllll|llll@{}}
\toprule
\textbf{Model}                                           & \multicolumn{4}{l|}{\textbf{No user context}}     & \multicolumn{4}{l}{\textbf{With user context}}    \\ \cmidrule(l){2-9} 
                                                         & \textbf{P} & \textbf{R} & \textbf{F} & \textbf{A} & \textbf{P} & \textbf{R} & \textbf{F} & \textbf{A} \\ \midrule
DeepSeek-r1 14b \cite{deepseek-ai_deepseek-r1_2025}       & 0.68       & 0.64       & 0.66       & 0.76       & 0.71       & 0.67       & 0.69       & 0.78       \\
DeepSeek-r1 32b \cite{deepseek-ai_deepseek-r1_2025}       & 0.74       & 0.54       & 0.62       & 0.76       & 0.77       & 0.59       & 0.67       & 0.79       \\
DeepSeek-r1 70b \cite{deepseek-ai_deepseek-r1_2025}       & 0.76       & 0.72       & 0.74       & 0.82       & 0.78       & 0.77       & 0.78       & 0.84       \\
Gemma 3 27b \cite{gemma_gemma_2025}                       & 0.64       & \textbf{0.92}       & 0.76       & 0.78       & 0.62       & \textbf{0.94}       & 0.75       & 0.77       \\
Gemma 4 26b \cite{gemma4team}                             & \textbf{0.92}       & 0.48       & 0.63       & 0.80       & \textbf{0.90}       & 0.44       & 0.59       & 0.78       \\
Gemma 4 31b \cite{gemma4team}                             & 0.87       & 0.78       & \textbf{0.82}       & \textbf{0.88}       & 0.86       & 0.80       & \textbf{0.83}       & \textbf{0.88}       \\
Gpt-oss 120b \cite{openai_gpt-oss-120b_2025}              & 0.74       & 0.83       & 0.78       & 0.83       & 0.74       & 0.86       & 0.79       & 0.84       \\
Gpt-oss 20b \cite{openai_gpt-oss-120b_2025}               & 0.69       & 0.66       & 0.68       & 0.77       & 0.69       & 0.66       & 0.68       & 0.77       \\
Llama 3.1 8b instruct \cite{grattafiori_llama_2024}       & 0.49       & 0.66       & 0.56       & 0.62       & 0.50       & 0.52       & 0.51       & 0.63       \\
Llama 3.3 70b \cite{grattafiori_llama_2024}               & 0.68       & 0.86       & 0.76       & 0.80       & 0.72       & 0.87       & 0.79       & 0.83       \\
Qwen 3 30b a3b instruct \cite{yang_qwen3_2025}            & 0.62       & 0.90       & 0.73       & 0.76       & 0.64       & 0.90       & 0.75       & 0.78       \\
Qwen 3 30b a3b thinking \cite{yang_qwen3_2025}            & 0.83       & 0.67       & 0.74       & 0.83       & 0.84       & 0.65       & 0.73       & 0.83       \\
Qwen 3 32b \cite{yang_qwen3_2025}                         & 0.76       & 0.78       & 0.77       & 0.83       & 0.76       & 0.86       & 0.81       & 0.85       \\
Qwen 3.6 27b \cite{qwenteamqwen35omnitechnicalreport2026} & 0.86       & 0.77       & 0.81       & 0.87       & 0.83       & 0.76       & 0.79       & 0.85       \\
Qwen 3.6 35b \cite{qwenteamqwen35omnitechnicalreport2026} & 0.85       & 0.73       & 0.78       & 0.85       & 0.85       & 0.71       & 0.78       & 0.85       \\ \bottomrule
\end{tabular}%
\caption{Text classification performance for binary antisemitism detection in two experiment setups---with and without user information as additional context (precision, recall, F1-score, and accuracy for the \textit{is\_antisemitic=True} minority class).}
\label{tab:performance_binary}
}
\end{table*}

With the expert-labeled data, we operationalize anti-Semitism detection as a multi-label text classification task that is performed via LLM-based in-context learning.

In a first step, we optimized the prompt to reduce common classification errors that we noticed during the rounds of sampling annotation candidates and manual labeling.
This involved mainly presenting further negative examples for each category to clarify their distinct meaning. Moreover, we developed two versions of this final prompt---one including author names and platform usernames in brackets as additional context, and one without user information. 
We assume that, given sufficient world knowledge, an LLM would be better able to predict whether an antisemitic utterance is neutral (descriptive) or critical if, in addition to the text, user information is provided. The weekly newspaper \textit{Jüdische Allgemeine}, for example, frequently reports on anti-Semitic incidents, but is highly unlikely to express anti-Semitic views itself.
The final codebook was presented to the LLM as the system prompt (see Appendix~\ref{app:online}). In included general rules for coding decisions as well as a short definition, key distinguishing features, and positive and negative examples for each category. The to-be-classified posts were presented as user prompts in batches of 10 examples, providing the post text and, optionally, the username/handle of the post author/channel in JSON format. The output was requested in JSON format as well, including the label decisions along with a comprehensive rationale for each label.

In a second step, we evaluated the classification performance of a wide range of open-weight LLMs to detect anti-Semitic categories. Tab.~\ref{tab:performance_binary} shows the results for the binary classification level, i.e., distinguishing antisemitic from non-antisemitic posts with and without providing user information as additional context.
We have recoded the LLM outputs into a binary variable that indicates whether the model detected at least one of the 26 categories.
The results show that automatic detection of anti-Semitic content can be achieved with medium-sized LLMs to a satisfactory extent. Large models with 70B and above parameters achieve F1-scores around 80~\%. The best performing models are the two dense `thinking' models 
Gemma 4 31b with an F1-score of 83~\% (88~\% accuracy), and Qwen 3 in the 32b parameter version with an F1-score of 81~\% (85~\% accuracy).
OpenAI's \textit{gpt-oss 120b} lags slightly behind with an F1-score of ca. 79~\% and 84~\% accuracy, but it classifies the validation dataset about five times faster than the newer, top-scoring models of the Gemma~4 and Qwen~3 model series. Providing user context, in general, helps most of the `older' LLMs to increase the precision and recall of anti-Semitic post detection. Interestingly, the effect vanishes for the most recent Qwen 3.6 and Gemma 4 models, which seem to be able to determine a framing, such as neutral reporting or critical commenting, on their own.

Distinguishing sub-categories of anti-Semitism is a much more difficult task. The performance evaluation in Tab.~\ref{tab:performance_subcategories} shows mixed results. 
About half of the classes achieve acceptable performances around 60~\% F1-score and higher. The other half, most of the smaller categories in the validation dataset, only achieve F1-scores below 50~\%. As the data for individual classes is extremely imbalanced, accuracy values are still very high, also for these classes. However, with precision scores even below 20~\% in five cases, there is a high risk of overestimating the prevalence for these categories with our automatic approach. When examining the rationales generated for the classification decisions, we find that smaller models, in particular, tend to hallucinate justifications for misclassifications that are not supported by the text of the post. Larger models and newer `thinking' models successfully take into account the constraints for labels specified by the codebook to determine when a particular category should be applied (and when it should not).

\begin{figure*}
    \centering
    \includegraphics[width=0.8\linewidth]{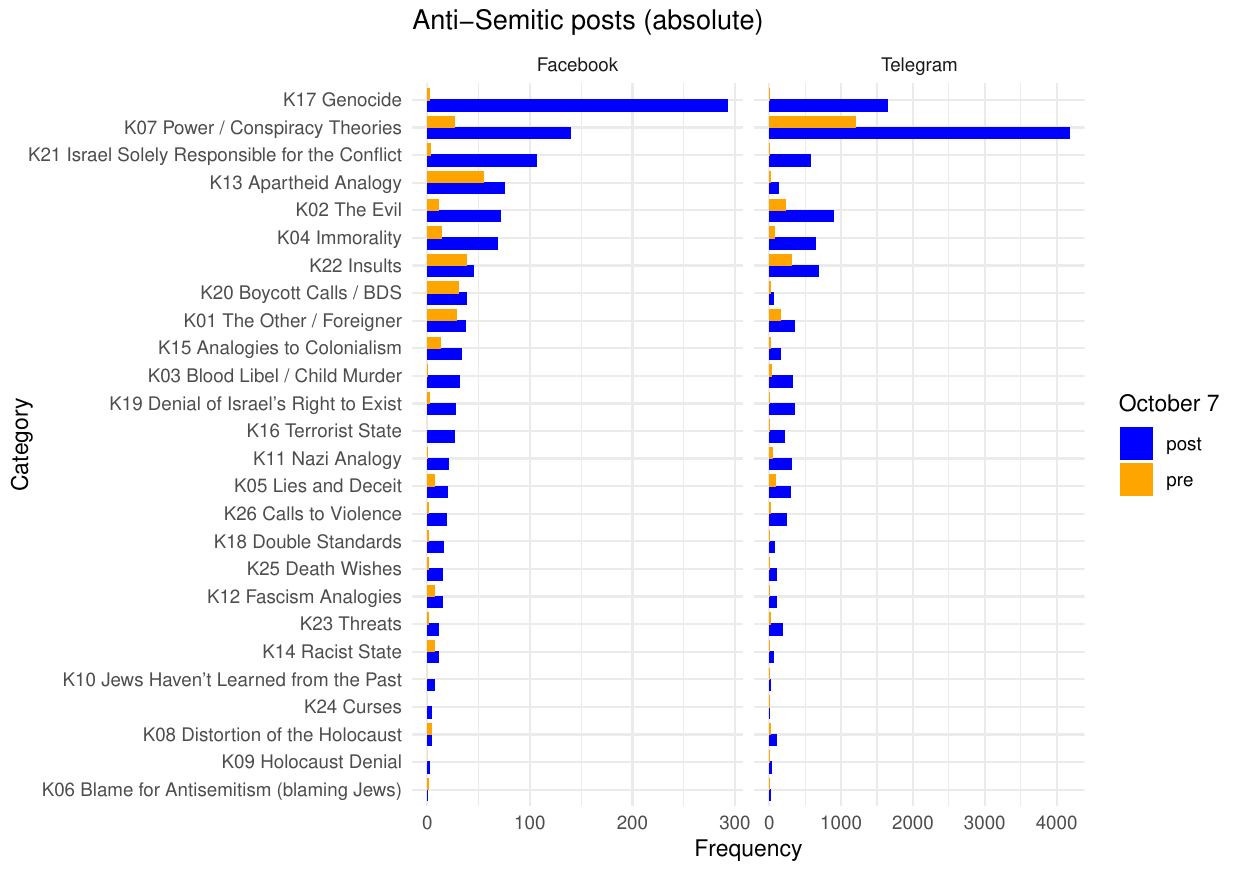}
    \caption{Frequency of anti-Semitic categories per platform}
    \label{fig:abs}
\end{figure*}

\begin{figure*}
    \centering
    \includegraphics[width=0.8\linewidth]{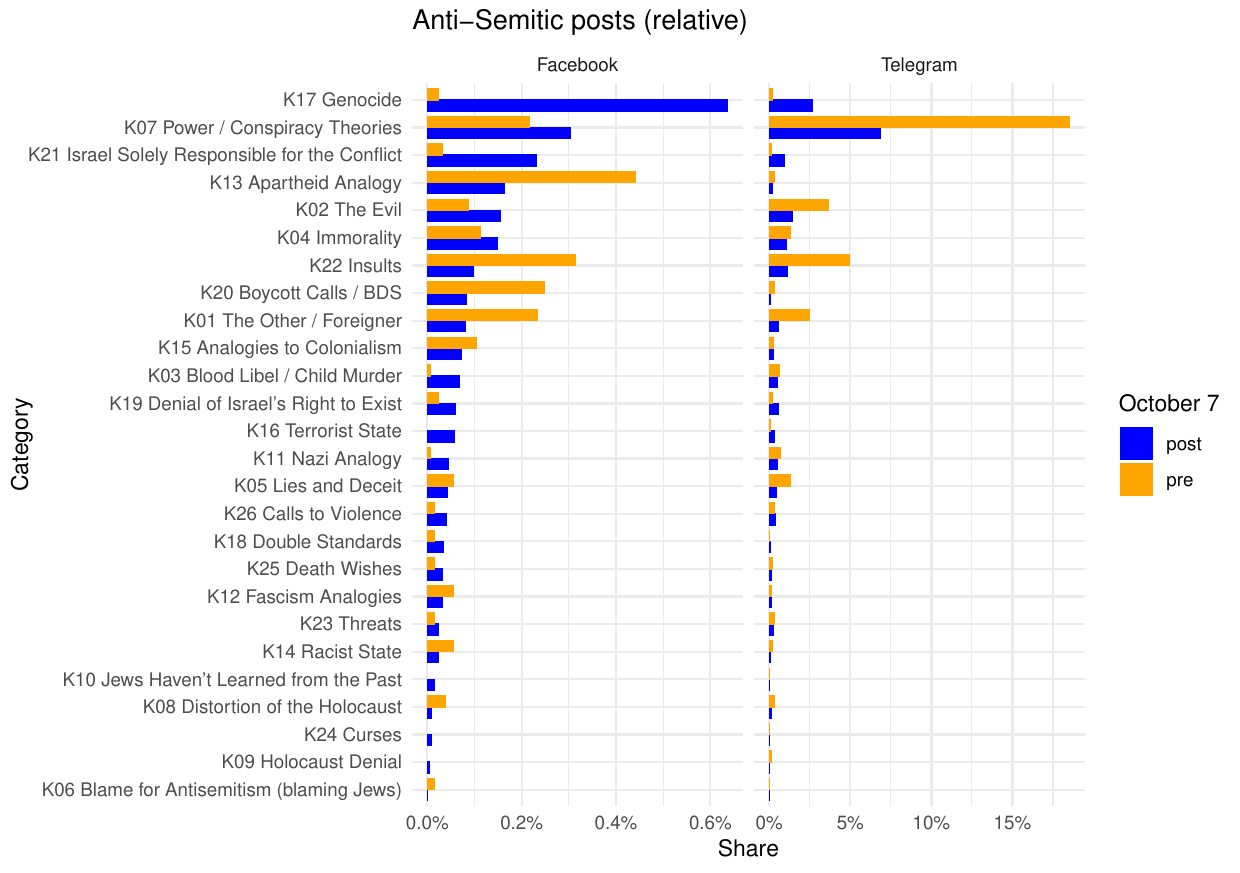}
    \caption{Share of anti-Semitic categories per platform}
    \label{fig:shares}
\end{figure*}

\section{Results}
\label{sec:results}

We used the optimized, final prompt together with \textit{gpt-oss 120b} to label the full dataset of posts from both platforms before and after the October 7 attacks.  
We chose this model for economic reasons as a compromise between validity and speed of the classification process.
Fig.~\ref{fig:abs} and Fig.~\ref{fig:shares} display the absolute and the relative number of anti-Semitic posts comparing the phase before and after the October 7 attack. In total, we observe about 10 times more anti-Semitic posts on Telegram compared to Facebook (1,028, i.e., about 1.7~\% for Facebook; 10,726, i.e., 15.9~\% for Telegram). 
In absolute numbers, anti-Semitic posts surged for all categories on both platforms. The numbers for individual categories, however, drastically differ.
On Facebook, accusations of genocide against Palestinians by the Israeli army (K17) dominate the post-terrorist attack phase, whereas previously it played a negligible role.\footnote{During the Gaza War, the intensity of Israel's actions led to intense debate about allegations of genocide. More and more observers take the view that the characteristics of genocide have been met in Gaza. However, we argue that this is probably not yet true for this study's period of investigation.} Telegram is dominated by classic anti-Semitic clichés of a Jewish conspiracy (K07) before and after October 7.  

While anti-Semitism increased across all categories, the picture gets more nuanced by looking at increases/decreases of shares among all posts related to Jewish/Israeli affairs on a platform (cf Fig.~\ref{fig:appinc}). Here, we can observe different patterns of the discursive development before and after the attacks. 
The genocide accusation shows the strongest increase in category share among all categories on Telegram. In general, the increases can be observed in categories of anti-Israel antisemitism (K17, K21, K19, K16, K18, K11). The classic anti-Semitic tropes of the powerful (K07), rootless (K01), evil (K02), or lying and deceitful (K05) Jew show the most significant relative decrease. On Facebook, in contrast, some classic tropes, such as evil (K02) and immorality (K04), show a noticeable relative increase. Other classic tropes, such as the accusation of blood libel and child murder (K03), are articulated in an updated form targeting Israel's military intervention in Gaza. Boycott calls (K20), insults (K22), and Apartheid analogies (K13) are relatively less articulated.

In summary, it can be observed that before the terrorist attack, there were clearly different patterns of use of anti-Semitic categories on both platforms, but these patterns have since become much more similar. The sharp increase in anti-Israel forms of anti-Semitism on Facebook has been accompanied by an increase in classic anti-Semitic statements. On Telegram, on the other hand, traditional anti-Semitism, which was already strong before, is now being reinforced by anti-Israel forms such that fringe and mainstream discourse now align more closely.

\begin{table}[ht]
\centering
{\fontsize{9.5}{11.5}\selectfont
\begin{tabular}{lrrrrr}
\toprule
\textbf{Class} & \textbf{P} & \textbf{R} & \textbf{F1} & \textbf{N} & \textbf{Acc.} \\
\midrule
K20 & 0.87 & 0.76 & 0.81 & 17 & 0.993 \\
K15 & 0.79 & 0.79 & 0.79 & 14 & 0.993 \\
K13 & 0.88 & 0.65 & 0.75 & 23 & 0.988 \\
K07 & 0.77 & 0.72 & 0.74 & 121 & 0.925 \\
K17 & 0.69 & 0.75 & 0.72 & 61 & 0.955 \\
K11 & 0.67 & 0.76 & 0.71 & 21 & 0.984 \\
K14 & 0.77 & 0.59 & 0.67 & 17 & 0.988 \\
K24 & 1.00 & 0.50 & 0.67 & 6 & 0.996 \\
K09 & 0.60 & 0.69 & 0.64 & 13 & 0.988 \\
K12 & 0.53 & 0.73 & 0.62 & 11 & 0.988 \\
K16 & 0.80 & 0.42 & 0.55 & 19 & 0.984 \\
K23 & 0.29 & 0.83 & 0.43 & 6 & 0.984 \\
K08 & 0.55 & 0.32 & 0.40 & 19 & 0.978 \\
K10 & 1.00 & 0.25 & 0.40 & 4 & 0.996 \\
K05 & 0.42 & 0.36 & 0.38 & 14 & 0.980 \\
K03 & 0.50 & 0.30 & 0.38 & 10 & 0.988 \\
K04 & 0.41 & 0.33 & 0.37 & 21 & 0.970 \\
K02 & 0.32 & 0.42 & 0.37 & 31 & 0.944 \\
K19 & 0.33 & 0.33 & 0.33 & 15 & 0.975 \\
K22 & 0.18 & 0.54 & 0.27 & 13 & 0.953 \\
K26 & 0.18 & 0.40 & 0.25 & 5 & 0.985 \\
K18 & 0.11 & 0.33 & 0.17 & 3 & 0.988 \\
K01 & 0.11 & 0.33 & 0.16 & 6 & 0.974 \\
K21 & 0.09 & 0.06 & 0.07 & 17 & 0.968 \\
K06 & 0.00 & 0.00 & 0.00 & 1 & 0.999 \\
\bottomrule
\end{tabular}
\caption{Text classification performance of gpt-oss 120b for individual categories (precision, recall, F1-score, and accuracy; N: number of `positive' test instances)}
\label{tab:performance_subcategories}
}
\end{table}

\section{Concluding remarks}

In this paper, we examined how the October 7, 2023, attacks affected antisemitic communication in German-language Facebook and Telegram posts. Relying on a tailored codebook with 26 categories of antisemitism and an expert-annotated validation set, we showed that medium- and large-scale open-weight LLMs can be used to automatically identify a broad spectrum of antisemitic tropes with sufficient reliability for downstream quantitative analysis. Applying a well-performing model to a comprehensive sample of Judaism- and Israel-related posts from both platforms, we found a substantial surge of antisemitism after October 7 on Facebook and Telegram alike, with Telegram exhibiting roughly an order of magnitude higher prevalence overall.

Our findings highlight that not only the amount but also the composition of antisemitic discourse shifted in the wake of the attacks. Before October 7, Telegram was dominated by classic antisemitic stereotypes centered on Jewish power and conspiracies, whereas Facebook showed comparatively lower levels of antisemitism overall. After the attacks, accusations of genocide in Gaza became the most salient category on Facebook, while the Telegram discourse combined pre-existing classic antisemitic tropes with a marked rise in Israel-focused accusations. In other words, the initially distinct antisemitic profiles of the two platforms began to converge, with anti-Israel antisemitism becoming more prominent on both, yet against different discursive baselines.

\section*{Limitations}

Methodologically, our study demonstrates both the promise and the current limits of LLM-based content analysis in highly sensitive domains. 
With our keyword-based data collection approach, we focused on posts related to Israeli and Jewish affairs. However, this method does not guarantee to include all anti-Semitic content that could be found on the platform.
Methodically, we show that carefully designed prompts, a conceptually grounded codebook, and expert-annotated validation data enable robust binary detection of antisemitic content and acceptable performance for a subset of more fine-grained categories. At the same time, the uneven distribution of categories, systematic confusions between semantically adjacent labels, and the context-dependence of antisemitic meaning point to the need for continued human oversight and iterative refinement. 
The task gets even more complicated, taking into consideration the ongoing academic and civil society debates about the appropriate definition of anti-Semitism and a constantly changing crisis in the Middle East that further polarizes the debate.
In this light, LLMs do not replace expert judgment. They rather function as scalability instruments that make large-scale empirical analyses of complex hate speech phenomena defined by experts at a certain point in time practically feasible.

\section*{Acknowledgments}

The authors disclose receipt of the following financial support for the research, authorship, and/or publication of this article: this article is based on research conducted in the DFG/FWF funded Research Unit 5656 'Communicative AI', project ‘P6 | Political discourse: ComAI and deliberative quality’ (funded by Deutsche Forschungsgemeinschaft (DFG, German Research Foundation) - 516511468).

\bibliography{custom}

\appendix

\section{Methodology}

\subsection{Subcategories}
\label{app:codebook}

Table~\ref{app:codebook} presents a summary of the codebook translated from German to English along with the number of annotated examples per category in the expert-annotated dataset.

\begin{table*}
\centering
\footnotesize
\begin{tabularx}{\textwidth}{lp{2.5cm}Xrr}
\toprule
\textbf{Code} & \textbf{Name (translation)} & \textbf{Short definition (translated from German)} & \textbf{N} & $\mathbf{\kappa}$\\
\midrule
K01 & The Other / Foreigner                        & Jews are constructed as fundamentally different or not belonging, a foreign out-group that does not really belong to the national or social community.                                   & 6   & 0.26 \\
K02 & Evil                                         & Attribution of an ontological, global evil nature to Jews, Israelis, Israel or Zionism, as if they had an inherently evil core or destructive essence.                                   & 31  & 0.38 \\
K03 & Blood libel / Child murder                   & Claims that Jews or Israelis deliberately kill non-Jewish children or use their blood, often framed as ritual or systematic killing without evidence.                                    & 10  & 0.57 \\
K04 & Immorality                                   & Jews or Israelis are portrayed as fundamentally without conscience or empathy, often combined with accusations of hypocrisy or instrumentalization of the Holocaust.                     & 21  & 0.32 \\
K05 & Lies and deceit                              & Jews or Israelis are depicted as inherently dishonest and manipulative, with generalized accusations of systematic lying, distortion and deception.                                      & 14  & 0.30 \\
K06 & Blaming Jews for antisemitism                & Responsibility for antisemitic incidents or hostility is shifted onto Jews themselves, reversing victim–perpetrator roles by claiming that Jewish behaviour causes antisemitism.         & 1   & 0.00 \\
K07 & Power and conspiracy narratives              & Attribution of exaggerated, often secret and global Jewish, Israeli or ``Zionist'' power that supposedly controls politics, media, economy or world events through conspiratorial plots. & 106 & 0.58 \\
K08 & Distortion of the Holocaust                  & Relativizing, shifting guilt or instrumentalized equations of the Holocaust, for example, downplaying its uniqueness or portraying perpetrators as less responsible.                      & 19  & 0.68 \\
K09 & Holocaust denial                             & Complete denial of core historical facts of the Holocaust, such as the existence of gas chambers or its character as a genocide, often demanding impossible ``proof''.                   & 13  & 0.51 \\
K10 & Jews have not learned from the past          & Moral failure is ascribed to Jews or Israelis by invoking their own historical suffering (especially the Holocaust) and claiming they have not ``learned the lesson''.                   & 4   & 0.50 \\
K11 & Nazi analogy                                 & Equating Israel, Israelis or Jews with Nazi Germany or neo-Nazis, including direct comparisons to Hitler, concentration camps or a ``second Holocaust''.                                 & 21  & 0.92 \\
K12 & Fascism analogies                            & Israel or Zionism is broadly labelled as fascist, equated with historical fascist regimes without careful differentiation, as a way to delegitimize the state.                           & 11  & 0.76 \\
K13 & Apartheid analogy                            & Generalized equation of Israel with Apartheid-era South Africa, often without contextual limits (e.g.\ to the West Bank) and used to frame Israel as an apartheid state.                 & 23  & 0.75 \\
K14 & Racist state                                 & Israel or Zionism is delegitimized as inherently racist, for example by claiming that a Jewish state is racist by definition or that Jewish supremacy is its core principle.             & 17  & 0.84 \\
K15 & Colonialism analogies                        & Israel is described as a colonial project of foreign (often European) settlers, portraying Jews as an alien colonial presence that ``stole'' the land.                                   & 14  & 0.75 \\
K16 & Terrorist state                              & Israel as a whole state is equated with a terrorist organization, for example by labelling it the ``biggest terrorist in the world'' or comparing it to ISIS or Al-Qaida.                & 19  & 0.91 \\
K17 & Genocide                                     & Israel, Israelis or Jews are accused of committing or pursuing genocide against Palestinians, often using the term ``genocide'' or ``Holocaust 2.0'' for current policies.               & 61  & 0.82 \\
K18 & Double standards                             & Israel, Israelis or Jews are singled out for exclusive moral condemnation or sanctions, while comparable violations by other states are ignored or downplayed.                           & 3   & 0.22 \\
K19 & Denial of Israel's right to exist            & Jewish statehood is rejected or the dissolution of Israel is demanded, including eliminatory slogans that negate Israel's legitimacy as a state.                                         & 15  & 0.63 \\
K20 & Boycott calls / BDS                          & Support for broad BDS-style campaigns targeting Israel as a state and society, aiming at its economic, cultural or academic isolation.                                                   & 17  & 0.55 \\
K21 & Israel's sole blame for the conflict         & A black-and-white narrative in which Israel or Israelis are portrayed as the sole cause of all violence, while the agency or responsibility of other actors is erased.                   & 17  & 0.23 \\
K22 & Insults                                      & Derogatory, dehumanizing or slur-like insults directed at Jews or Israelis because of their (perceived) Jewish or Israeli identity.                                                      & 13  & 0.56 \\
K23 & Threats                                      & Threats of future violence or punishment against Jews or Israelis, often using formulations about coming retribution or that ``their time will come''.                                   & 6   & 0.59 \\
K24 & Curses                                       & Wishes for divine, religious or karmic punishment of Jewish or Israeli targets, invoking God's curse or fate to bring suffering upon them.                                               & 6   & 0.00 \\
K25 & Death wishes                                 & Explicit or implicit wishes for the death or annihilation of Jews, Israelis, Israel or supporters of Israel, including references to finishing what the Holocaust started.               & n/a & 0.00 \\
K26 & Endorsement / incitement / wish for violence & Praise, justification or calls for violence against Jewish or Israeli targets, including support for attacks and appeals to continue or escalate violent struggle.                       & 5   & 0.26 \\
\bottomrule
\end{tabularx}
\caption{Categories of antisemitism, short definitions, number of examples, and Cohen's Kappa inter-coder agreement per category in the expert validation set. Names and descriptions are translated from the German original data. For the full, original codebook and corresponding LLM prompts, see the online resources (Appendix~\ref{app:online}).}
\label{tab:codebook}
\end{table*}

\subsection{Label co-occurrence}
\label{app:cooc}

\begin{figure*}
    \centering
    \includegraphics[width=0.7\linewidth]{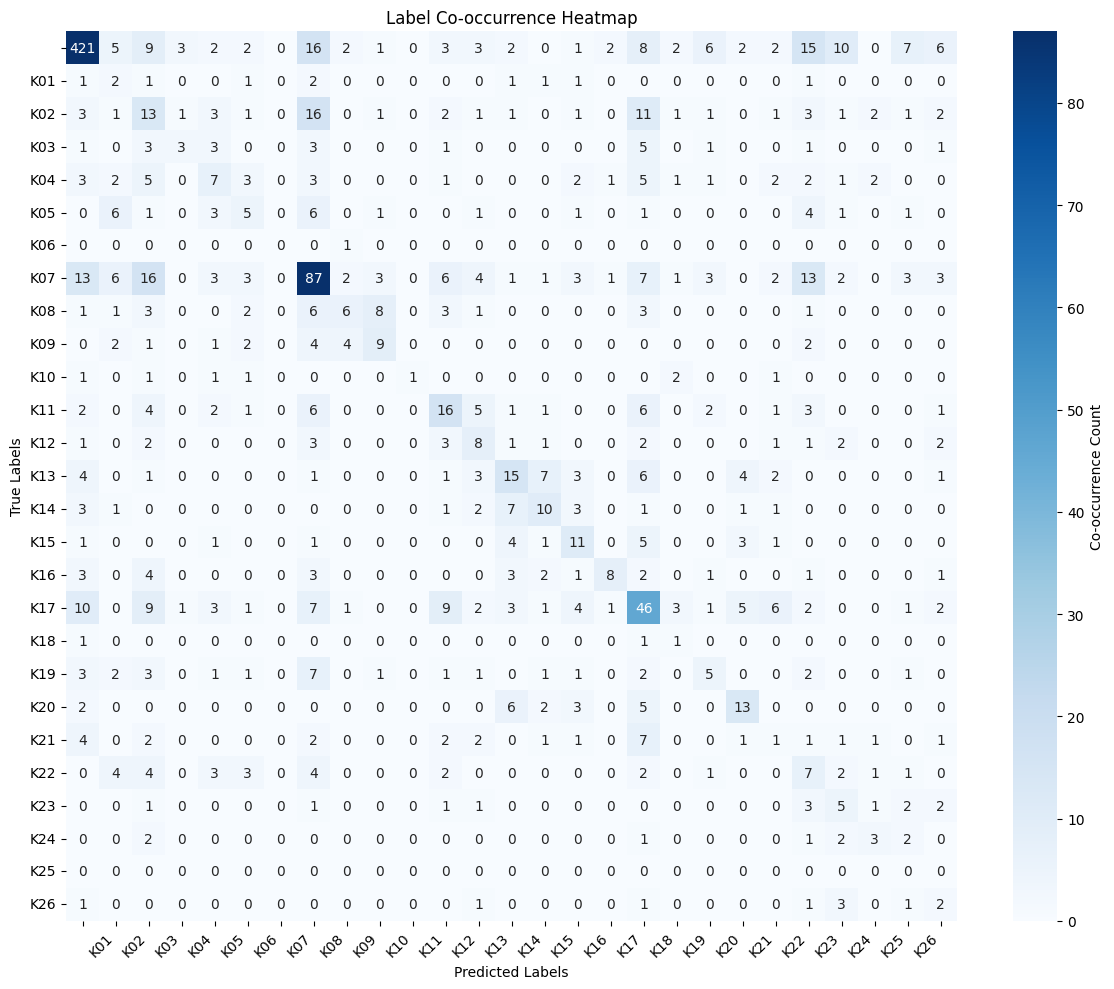}
    \caption{Label co-occurrence heatmap of true (rows) and predicted (columns) labels (co-occurrence counts show which true labels tend to get predicted together).}
    \label{fig:heatmap}
\end{figure*}

Figure ~\ref{fig:heatmap} displays co-occurrence counts of true labels and predicted labels with our best-performing model setup on the validation dataset. It reveals which categories are semantically close and, thus, get frequently confused during automatic coding. For instance, the category `Evil' (K02) is frequently mixed up with secret power and conspiracies (K07) and accusations of conducting a genocide in Gaza (K17).

\subsection{Increase/decrease of category shares}
\begin{figure*}
    \centering
    \includegraphics[width=0.48\linewidth]{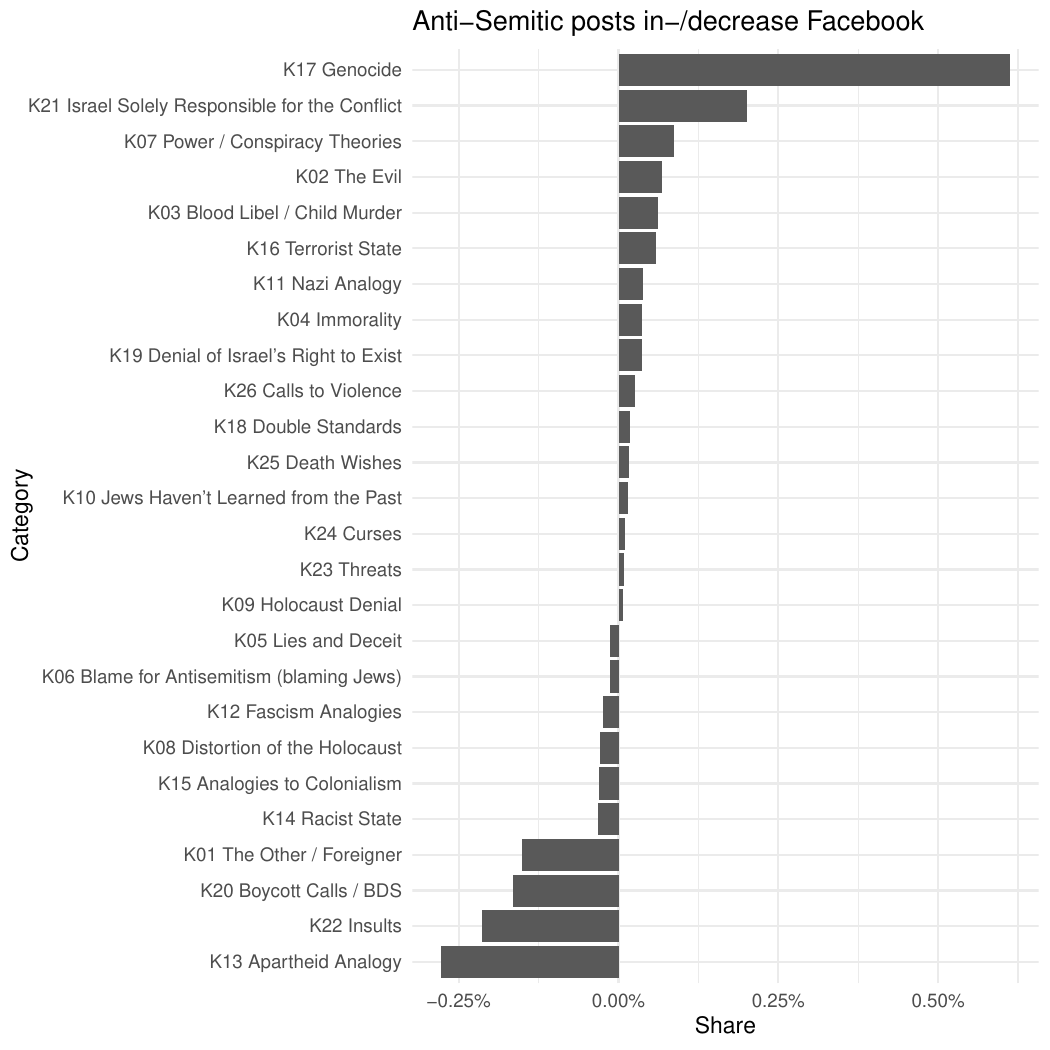}\hspace{1em}
    \includegraphics[width=0.48\linewidth]{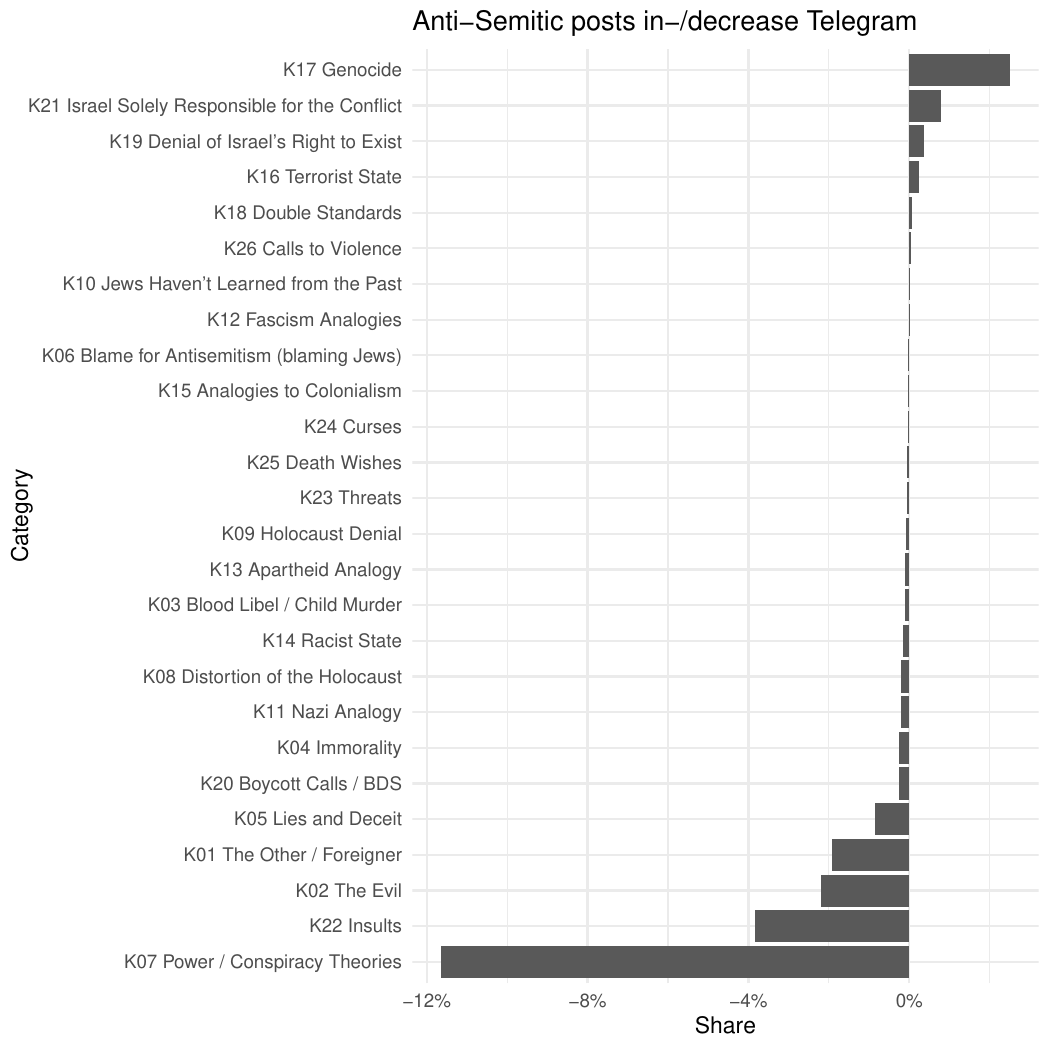}
    \caption{Difference of category shares pre/post the October 7 attacks with regard to the overall number of posts on the platform.}
    \label{fig:appinc}
\end{figure*}

Figure~\ref{fig:appinc} shows the different developments of anti-Semitic category shares that lead to an annealing of the discourse patterns on both platforms.

\section{Online Resources}
\label{app:online}

See the Zenodo repository published along with this paper to obtain 
\begin{itemize}
    \item the final system prompt used to classify anti-Semitism categories with LLMs
    \item the expert annotated validation set
\end{itemize}

\vspace{.5em}
\noindent
\href{https://doi.org/10.5281/zenodo.21420613}{https://doi.org/10.5281/zenodo.21420613}

\end{document}